\documentclass[12pt,preprint]{aastex}

\slugcomment{Accepted by ApJS}

\shorttitle{Equation of State in a RHD Code}
\shortauthors{Choi \& Wiita}

\begin{document}

\title{A Multidimensional Relativistic Hydrodynamics Code with a General
       Equation of State}

\author{Eunwoo Choi\altaffilmark{1} and Paul J. Wiita\altaffilmark{2,3}}

\altaffiltext{1}{Department of Astronomy and Atmospheric Sciences,
                 Kyungpook National University, Daegu 702-701, Korea;
                 echoi@knu.ac.kr}
\altaffiltext{2}{Department of Physics and Astronomy, Georgia State
                 University, P.O. Box 4106, Atlanta, GA 30302-4106, USA}
\altaffiltext{3}{Department of Physics, The College of New Jersey,
                 P.O. Box 7718, Ewing, NJ 08628, USA}

\begin{abstract}

The ideal gas equation of state with a constant adiabatic index, although
commonly used in relativistic hydrodynamics, is a poor approximation
for most relativistic astrophysical flows.
Here we propose a new general equation of state for a multi-component
relativistic gas which is consistent with the Synge equation of state
for a relativistic perfect gas and is suitable for
numerical (special) relativistic hydrodynamics.
We also present a multidimensional relativistic hydrodynamics code
incorporating the proposed general equation of state, based on the HLL
scheme, which does not make use of a full characteristic decomposition of
the relativistic hydrodynamic equations.
The accuracy and robustness of this code is demonstrated in
multidimensional calculations through several highly relativistic test
problems taking into account nonvanishing tangential velocities.
Results from three-dimensional simulations of relativistic jets
show that the morphology and dynamics of the relativistic jets are
significantly influenced by the different equation of state and by
different compositions of relativistic perfect gases.
Our new numerical code, combined with our proposed equation of state
is very efficient and robust, and unlike previous codes, it gives
very accurate results for thermodynamic variables in relativistic
astrophysical flows.

\end{abstract}


\section{Introduction}

Many high energy astrophysical phenomena, including accretion flows, jet
flows, gamma-ray bursts, and pulsar winds involve relativistic flows.
In powerful extragalactic radio sources, for example, ejections from
galactic nuclei produce intrinsic beam Lorentz factors of usually more than $5$
and apparently up to $\sim50$,
which are required to explain the apparent superluminal
motions observed in extragalactic radio sources associated with active
galactic nuclei \citep[e.g.,][]{lis09}.
In the expansion of many relativistic jets the internal thermal energy of a gas is
converted into bulk kinetic energy so as to reach a high Lorentz factor in
a short distance.
Then this kinetic energy is dissipated by shock
interactions, mostly by terminal shock complexes, and partially by
internal shocks within the jets as they
propagate over long distances \citep[e.g.,][]{nor82}.
Since relativistic flows are inherently nonlinear and complex, in
addition to possessing large Lorentz factors, numerical simulations have
been performed to investigate such relativistic flows, for example, in
the propagation of relativistic extragalactic jets
\citep[e.g.,][]{dun94,mar97,ros99,hug02,cho07}.

Many explicit finite difference
schemes originally applied to classical hydrodynamics have been employed
to treat special relativistic hydrodynamics numerically.
These schemes to solve the relativistic hydrodynamic equations are based
on either exact or approximate solutions to the local Riemann problem
\citep{sch93,fal96,don98,alo99,del02,ann03,mig05a,mig05b}.
Some of these schemes adopt local characteristic decomposition of
the Jacobian matrix to build numerical fluxes.
It is often difficult to build characteristic decomposition in some
regimes, especially in ultrarelativistic limits, due to the degeneracy of
the characteristic information.
Thus the use of an alternative scheme becomes sensible when the
characteristic decomposition is unknown.
The HLL scheme proposed by \citet{har83} for classical hydrodynamics is
based on an approximate Riemann solver that does not require full,
and numerically expensive, characteristic decomposition.
This feature of the HLL scheme makes its use very attractive, especially
in multidimensions, where computational efficiency and robustness is
extremely important.
This scheme was applied first to relativistic hydrodynamics by
\citet{sch93} in one dimension and by \citet{dun94} in multidimensions.

It is worth stressing that many treatments of relativistic astrophysical problems
have assumed a ideal gas equation of state with a constant polytropic
index, but this is a reasonable approximation only if the gas is either
strictly subrelativistic or ultrarelativistic.
However, when the gas is semirelativistic or when the gas has two
components, e.g., nonrelativistic protons and relativistic
electrons, this assumption is no longer correct.
This was shown for the relativistic perfect gas law by \citet{syn57},
where the exact form of an equation of state relating thermodynamic
quantities of specific enthalpy and temperature is completely described
in terms of modified Bessel functions.

Since the correct equation of state for the relativistic perfect gas has been
recognized as being important, several investigations
with a more general equation of state have been reported in numerical
relativistic hydrodynamics.
\citet{fal96} described an upwind numerical code for special relativistic
hydrodynamics with the Synge equation of state for multi-component
relativistic gas.
More recently, \citet{mig05b} used, in their upwind relativistic
hydrodynamic code, a simple equation of state that closely approximates
the Synge equation of state for a single-component relativistic gas.
Several numerical simulations in the context of relativistic extragalactic
jets make use of the general equation of state to account for transitions
from nonrelativistic to relativistic temperature
\citep{kom98,sch02,per07,mel08,ros08}.
\citet{sch02} used the Synge equation of state for different compositions,
including pure leptonic and baryonic plasmas, to investigate the influence
of the composition of relativistic extragalactic jets on their long-term
evolution.
Similarly, \citet{mel08} studied the relativistic extragalactic jet
deceleration through density discontinuities by using the Synge-like
equation of state with a variable polytropic index.

In this work we propose an analytical form of equation of state for the
multi-component relativistic perfect gas that is consistent with the
Synge equation of state in the relativistic regime.
This proposed equation of state is suitable for a
numerical code from the computational point of view, unlike the Synge
equation of state, which involves the computation of Bessel functions.
We then build a multidimensional relativistic hydrodynamics code based
on the HLL scheme using the proposed equation of state, and demonstrate
the accuracy and robustness of this code by presenting several test problems
and numerical simulations.
In particular, we plan to use this code to simulate relativistic
extragalactic jets that are probably composed of a mixture of relativistic
particles of different masses.
Numerical simulations of relativistic jets of different compositions are
challenging, but are made tractable by using the proposed equation of state
that accounts for different compositions of relativistic gas.

This paper is organized as follows.
In \S 2 we present the equations of motion and the Synge equation of
state, and we describe the proposed general equation of state for the
relativistic gas.
In \S 3 we describe a relativistic hydrodynamics code based on the HLL
scheme, incorporating this proposed equation of state.
In \S\S 4 and 5 we present numerical tests and simulations with the code
to demonstrate the performance of the code.
A conclusion is given in \S 6.

\section{Relativistic Hydrodynamics}

\subsection{Equations of Motion}

The motion of relativistic gas is described by a system of conservation
equations.
The equations in special relativistic hydrodynamics are written in a
covariant form \citep{lan59,wil03} as
\begin{equation}
\partial_\alpha\left(\rho U^\alpha\right) = 0,
\end{equation}
\begin{equation}
\partial_\alpha\left(\rho h U^\alpha U^\beta+p g^{\alpha\beta}\right) = 0.
\end{equation}
Here, $\partial_\alpha = \partial/\partial x^\alpha$ is the covariant
derivative with respect to spacetime coordinates $x^\alpha = [t,x_j]$,
$U^\alpha = [\Gamma,\Gamma v_j]$ is the normalized ($U^\alpha U_\alpha = -1$)
four-velocity vector, where $\Gamma$ is the Lorentz
factor, and $g^{\alpha\beta} = \mathrm{diag}\{-1,1,1,1\}$ is the metric
tensor in Minkowski space.
The speed of light is set to unity ($c = 1$) in this work.
Greek indices (e.g., $\alpha$, $\beta$) denote the spacetime components
while Latin indices (e.g., $i$, $j$) indicate the spatial components.
The rest mass density, specific enthalpy, and pressure in the local rest
frame are denoted by $\rho$, $h$, and $p$, respectively.

In Cartesian coordinates these relativistic hydrodynamic equations can
be written in conservative form as
\begin{equation}
\frac{\partial\mbox{\boldmath$q$}}{\partial t}+\frac{\partial\mbox{\boldmath$F$}_x}{\partial x}
+\frac{\partial\mbox{\boldmath$F$}_y}{\partial y}+\frac{\partial\mbox{\boldmath$F$}_z}{\partial z} = 0,
\end{equation}
where \mbox{\boldmath$q$} is the state vector of conservative variables
and $\mbox{\boldmath$F$}_x$, $\mbox{\boldmath$F$}_y$, and
$\mbox{\boldmath$F$}_z$ are respectively the flux vectors in the $x$,
$y$, and $z$-directions, defined by
\begin{equation}
\mbox{\boldmath$q$} = \left[\matrix{D\cr M_x\cr M_y\cr M_z\cr E}\right],~~
\mbox{\boldmath$F$}_x = \left[\matrix{D v_x\cr M_x v_x+p\cr M_y v_x\cr M_z v_x\cr\left(E+p\right)v_x}\right].
\end{equation}
The flux vectors $\mbox{\boldmath$F$}_y$ and $\mbox{\boldmath$F$}_z$ are
given by properly permuting indices.
The conservative variables $D$, $M_x$, $M_y$, $M_z$, and $E$ represent
respectively the mass density, three components of momentum density, and
energy density in the reference frame.

The variables in the reference frame are nonlinearly coupled to those in
the local rest frame via the transformations
\begin{equation}
D = \Gamma\rho,
\end{equation}
\begin{equation}
M_x = \Gamma^2\rho h v_x,~M_y = \Gamma^2\rho h v_y,~M_z = \Gamma^2\rho h v_z,
\end{equation}
\begin{equation}
E = \Gamma^2\rho h-p,
\end{equation}
where the Lorentz factor is given by $\Gamma = 1/\sqrt{1-v^2}$ with
$v^2 = v_x^2+v_y^2+v_z^2$.

\subsection{Equation of State}

The system of conservation equations describing the motion of
relativistic gas is completed with an equation of state that relates
the thermodynamic quantities of specific enthalpy, rest mass density, and
pressure.
In general the equation of state can be expressed with the specific
enthalpy expressed as a function of the rest mass density and the pressure
\begin{equation}
h = h(\rho,p).
\end{equation}
The sound speed $c_s$ is then defined as
\begin{equation}
c_s^2 = \frac{\rho}{h}\frac{\partial h}{\partial\rho}\left(1-\rho\frac{\partial h}{\partial p}\right)^{-1}.
\end{equation}
The explicit form of the sound speed depends on the particular choice of
the equation of state.

The exact form of equation of state for a relativistic perfect gas
composed of multiple components was derived by \citet{syn57}.
For a single-component relativistic gas the equation of state is written as
\begin{equation}
h = \frac{K_3(\xi)}{K_2(\xi)},
\end{equation}
where $K_2$ and $K_3$ are respectively the modified Bessel function of
the orders two and three and $\xi = \rho/p$ is a measure of inverse temperature.
Under the Synge equation of state a relativistic perfect gas is
entirely described in terms of the modified Bessel functions.
The Synge equation of state for a relativistic gas can be written
in the form
\begin{equation}
h = 1+\frac{\gamma_r^\ast}{\gamma_r^\ast-1}\frac{p}{\rho},
\end{equation}
where the quantity $\gamma_r^\ast$ is defined by
\begin{equation}
\gamma_r^\ast = \frac{h-1}{h-1-\xi^{-1}}.
\end{equation}
Then the sound speed is written as
\begin{equation}
c_s^2 = \frac{\gamma_r p}{\rho h},
\end{equation}
and the relativistic adiabatic index $\gamma_r$ is given by
\begin{equation}
\gamma_r = \frac{h^\prime\xi^2}{h^\prime\xi^2+1},
\end{equation}
where $h^\prime = dh/d\xi$.
The quantities $\gamma_r^\ast$ and $\gamma_r$ are constant and equal if
the gas remains ultrarelativistic or subrelativistic
(i.e., $\gamma_r^\ast = \gamma_r = 4/3$ for $\xi\ll1$ or $5/3$ for $\xi\gg1$).
For the intermediate regime $\gamma_r^\ast$ and $\gamma_r$ vary slightly
differently between the two limiting cases \citep{fal96}.

For a multi-component relativistic gas, the direct use of the Synge
equation of state involves the computation of Bessel functions, thus requires
significant computation cost and results in computational inefficiency.
Here we propose a new general equation of state for multi-component
relativistic gas that uses analytical expression and is more efficient
and suitable for numerical computations.
We suppose that the relativistic gas is composed of electrons,
positrons, and protons although more components of relativistic gas
easily can be considered.
The total number density $n$ is then given by
\begin{equation}
n = \sum n_i = n_{e^-}+n_{e^+}+n_{p^+},
\end{equation}
where $n_{e^-}$, $n_{e^+}$, and $n_{p^+}$ are the electron, positron,
and proton number densities, respectively.
We ignore the production or annihilation of electron-positron pairs and
assume the composition of electrons, positrons, and protons is
maintained.
The assumption of charge neutrality gives us the relations
$n_{e^-} = n_{e^+}+n_{p^+}$ and $n = 2n_{e^-}$.
The total rest mass density and pressure are respectively given by
$\rho = \sum\rho_i = \sum n_i m_i$ and $p = \sum p_i = \sum n_i k T$,
where $k$ is the Boltzmann constant and $T$ is the temperature.

For our equation of state for multi-component relativistic gas we adopt
the equation of state, previously introduced by \citet{mat71} and later
used by \citet{mel04}, which closely reproduces the Synge equation of
state for a single-component relativistic gas \citep{mig05b}.
The equation of state takes the form
\begin{equation}
\frac{p}{\rho} = \frac{1}{3}\left(\frac{e}{\rho}-\frac{\rho}{e}\right),
\end{equation}
where $e$ is the energy density in the local rest frame.
It can be solved for the specific enthalpy using $\rho h = e+p$ as
\begin{equation}
h = \frac{5}{2}\frac{p}{\rho}+\sqrt{\frac{9}{4}\left(\frac{p}{\rho}\right)^2+1}.
\end{equation}
For a multi-component relativistic gas, the total enthalpy is then given by
$\rho h = \sum\rho_i h_i = \sum[(5/2)n_i k T+\sqrt{(9/4)(n_i k T)^2+(n_i m_i)^2}]$.
After we express the total enthalpy in terms of each component, we can
eliminate $n_{e^+}$ using the above relations given in the charge
neutrality assumption.
Our proposed equation of state is obtained by simplifying the expression
of the total enthalpy.
We find the resulting equation of state for a multi-component relativistic gas
to be
\begin{equation}
h = \frac{5}{2}\frac{1}{\xi}
+\left(2-\chi\right)\left[\frac{9}{16}\frac{1}{\xi^2}+\frac{1}{\left(2-\chi+\chi\mu\right)^2}\right]^{1/2}
+\chi\left[\frac{9}{16}\frac{1}{\xi^2}+\frac{\mu^2}{\left(2-\chi+\chi\mu\right)^2}\right]^{1/2},
\end{equation}
where $\chi = n_{p^+}/n_{e^-}$ is the relative fraction of proton and
electron number densities and $\mu = m_p/m_e$ is the mass ratio of
proton to electron.
We note that $\chi = 0$ represents an electron-positron gas while $\chi = 1$
indicates an electron-proton gas.
The proposed equation of state reduces to the equation of state in
equation (17) when the electron-positron gas ($\chi = 0$) is considered.
This proposed equation of state holds only in the limit where
the composition of the plasma is fixed in space and time.
In reality, however, the plasma composition may change through fluid
mixing or the creation or annihilation of electron-positron pairs,
either of which can alter the equation of state by effectively changing
$\chi$ in different regions of the fluid.
Nonetheless, this assumption of
constant $\chi$ is a useful first order approximation in many situations.

\begin{figure}
\begin{center}
\includegraphics[scale=0.8]{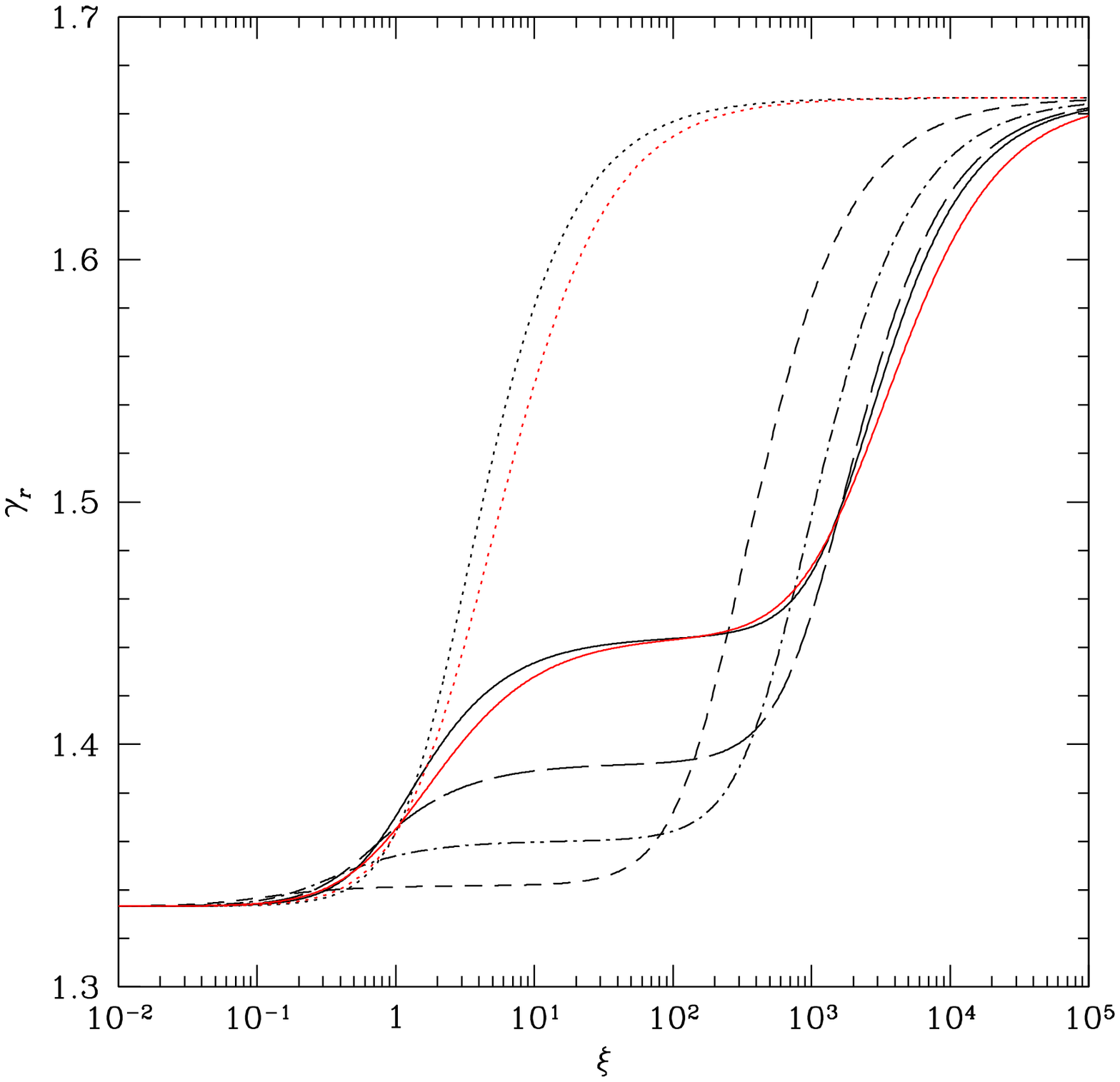}
\end{center}
\caption{The relativistic adiabatic index, $\gamma_r$, as function of
inverse temperature, $\xi$, for different compositions of relativistic
gas.
The compositions of $\chi = 0$ (electron-positron), $0.1$, $0.3$, $0.6$,
and $1$ (electron-proton) are shown using dotted, short dashed,
dot-short dashed, long dashed, and solid curves, respectively.
The exact Synge solutions for electron-positron and electron-proton gases
are respectively drawn as the red dotted and solid lines for comparison.}
\label{fig1}
\end{figure}

Using the proposed equation of state we show in Figure \ref{fig1} the
relativistic adiabatic indices, $\gamma_r$,
as functions of inverse temperature $\xi$, for several different compositions
of the relativistic gas.
Compositions of $\chi = 0$, $0.1$, $0.3$, $0.6$, and $1$ are shown.
The value of $\gamma_r$ critically depends on
the composition of the relativistic gas as protons become
relativistic at a much higher temperature than do electrons.
Different proton fractions cause the adiabatic index to vary substantially at
intermediate temperatures.
For electron-positron and electron-proton gases, we compare $\gamma_r$
values from our equation of state with $\gamma$ values from the Synge
equation of state as given in Figure 1 of \citet{fal96}.
Direct measurements of relative errors give
$|\gamma_r-\gamma|/\gamma\lesssim0.4\%$ for an electron-positron gas and
$\lesssim0.3\%$ for an electron-proton gas.

\begin{figure}
\begin{center}
\includegraphics[scale=0.8]{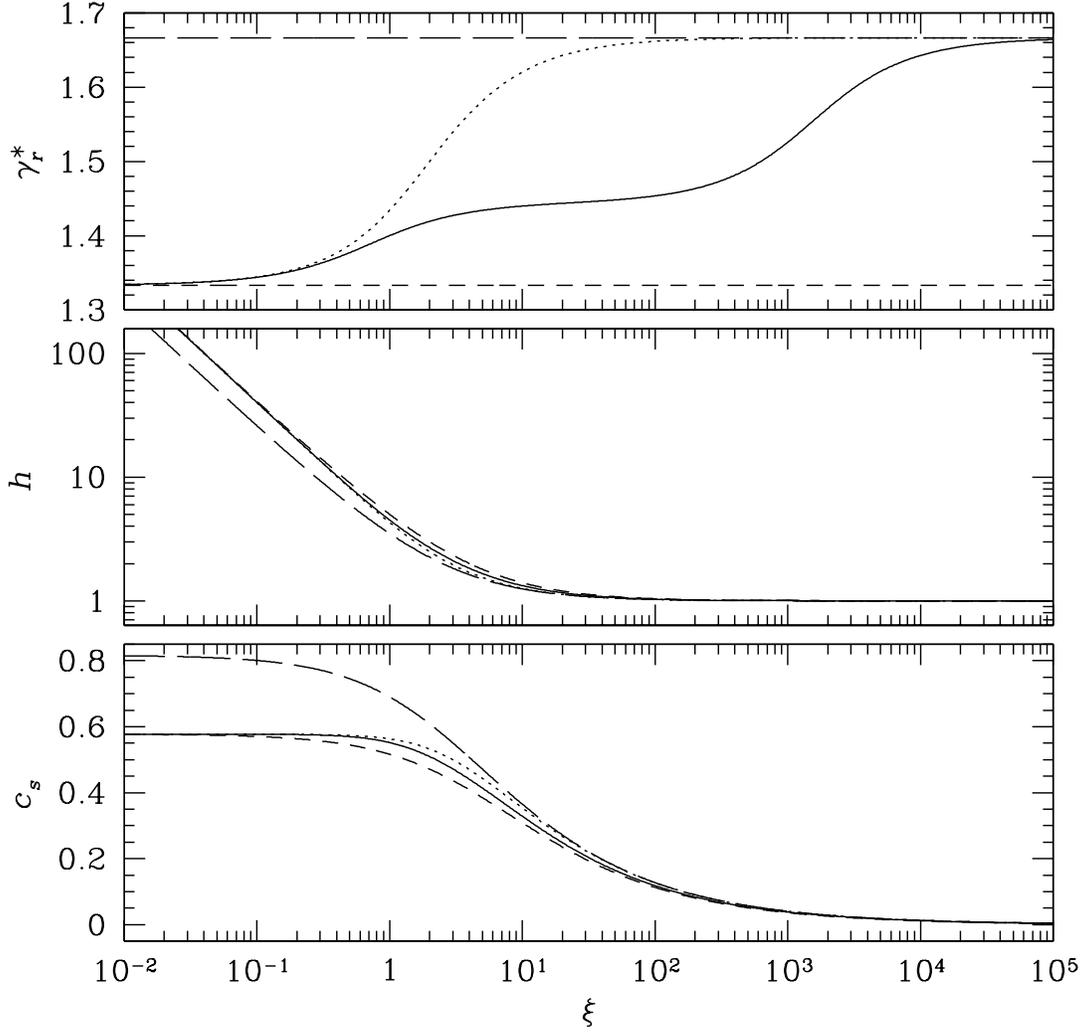}
\end{center}
\caption{The quantity $\gamma_r^\ast$, specific enthalpy $h$, and sound
speed $c_s$, as functions of inverse temperature $\xi$ for the different
equation of state.
The long dashed and short dashed lines correspond to the
ideal gas equation of state with constant adiabatic index $\gamma = 5/3$
and $4/3$, respectively.
Results for the proposed general equation of state for electron-positron
and electron-proton gases are shown by the dotted and solid lines, respectively.}
\label{fig2}
\end{figure}

Figure \ref{fig2} shows the quantities $\gamma_r^\ast$, specific enthalpy
$h$, and sound speed $c_s$, as functions of inverse temperature $\xi$, for
different equations of state.
Results are shown for the ideal gas equation of state with fixed
adiabatic indices of $\gamma = 5/3$
and $4/3$ as well as for the proposed general equation of state for pure
electron-positron and pure electron-proton gases.
The thermodynamic quantities computed using the proposed general equation
of state asymptotically approach those computed using the ideal gas equation
of state with $\gamma = 4/3$ in the hot gas limit ($\xi\ll1$) and
$\gamma = 5/3$ in the cold gas limit ($\xi\gg1$).
For intermediate regimes the thermodynamic quantities vary between those
two limiting cases, depending on the composition of the relativistic gas.

\section{Numerical Scheme}

We now describe a relativistic hydrodynamics code incorporating the
proposed equation of state for a relativistic perfect gas, based on the
HLL scheme originally proposed by \citet{har83} for classical
hydrodynamics.
The HLL scheme avoids a full characteristic decomposition of the
relativistic hydrodynamic equations and uses an approximate solution to
the Riemann problem where the two constant states separated by the
middle contact wave are averaged into a single intermediate state.
The HLL scheme requires accurate estimates of the maximum and minimum
wave speeds for the solution of the Riemann problem.
In classical hydrodynamics \citet{ein88} proposed good ways to estimate
the wave speeds based on the maximum and minimum eigenvalues of the
Jacobian matrix.

The wave speeds needed in our formulation for relativistic flows can be
similarly estimated from the maximum and minimum eigenvalues $a_x^\pm$ of the
Jacobian matrix,
$A_x = \partial\mbox{\boldmath$F$}_x(\mbox{\boldmath$u$})/\partial\mbox{\boldmath$q$}(\mbox{\boldmath$u$})$,
where $\mbox{\boldmath$u$} = [\rho,v_x,v_y,v_z,p]^\mathrm{T}$ is the
state vector of primitive variables, and
\begin{equation}
a_x^\pm = \frac{\left(1-c_s^2\right)v_x\pm\sqrt{\left(1-v^2\right)c_s^2
\left[1-v^2c_s^2-\left(1-c_s^2\right)v_x^2\right]}}{1-v^2c_s^2}.
\end{equation}
The maximum and minimum eigenvalues are based on a simple application of
the relativistic addition of velocity components decomposed into
coordinates directions and simply reduce to
$a_x^\pm = (v_x\pm c_s)/(1\pm v_x c_s)$ in the one-dimensional case.

Numerical integration of relativistic hydrodynamic equations advances by
evolving the state vector of conservative variables in time.
However, in order to compute the flux vectors for the evolution, the
primitive variables \mbox{\boldmath$u$} involved in the flux vectors
should be recovered from the conservative variables \mbox{\boldmath$q$}
at each time step by the inverse transformation
\begin{equation}
\rho = \frac{D}{\Gamma},
\end{equation}
\begin{equation}
v_x = \frac{M_x}{E+p},~v_y = \frac{M_y}{E+p},~v_z = \frac{M_z}{E+p},
\end{equation}
\begin{equation}
p = \Gamma D h-E.
\end{equation}
The inverse transformation is nonlinearly coupled and reduces to a single
equation for the pressure
\begin{equation}
f(p) = \Gamma(p)D h(\xi(p))-E-p = 0.
\end{equation}
This nonlinear equation can be solved numerically using a Newton-Raphson
iterative method in which the derivative of the equation with respect to
pressure is given by
\begin{equation}
\frac{d f}{d p} = \frac{d\Gamma}{d p}D h+\Gamma D\frac{d h}{d\xi}\frac{d\xi}{d p}-1.
\end{equation}
Once the pressure is found numerically, the rest mass density and
velocity are recovered by the inverse transformation.
This procedure of inversion from conservative to primitive variables is
valid for a general equation of state by specifying the expression of
specific enthalpy.

The numerical integration of relativistic hydrodynamic equations proceeds
on spatially discrete numerical cells in time, based on the finite
difference method.
In our implementation the state vector $\mbox{\boldmath$q$}_i^n$ at the
cell center $i$ at the time step $n$ is updated by calculating the
numerical flux vector $\mbox{\boldmath$f$}_{x,i+1/2}^{n+1/2}$ along the
$x$-direction at the cell interface $i+1/2$ at the half time step $n+1/2$ as follows
\begin{equation}
\mbox{\boldmath$q$}_i^{n+1} = \mbox{\boldmath$q$}_i^n-\frac{\Delta t^n}{\Delta x}
\left(\mbox{\boldmath$f$}_{x,i+1/2}^{n+1/2}-\mbox{\boldmath$f$}_{x,i-1/2}^{n+1/2}\right).
\end{equation}
The numerical flux vector is calculated from the approximate Riemann
solution and is given in the form
\begin{eqnarray}
\mbox{\boldmath$f$}_{x,i+1/2}^{n+1/2} &=& \frac{a_{x,i+1/2}^+\mbox{\boldmath$F$}_x(\mbox{\boldmath$u$}_{L,i+1/2}^{n+1/2})
-a_{x,i+1/2}^-\mbox{\boldmath$F$}_x(\mbox{\boldmath$u$}_{R,i+1/2}^{n+1/2})}{a_{x,i+1/2}^+-a_{x,i+1/2}^-} \nonumber \\
&-&\frac{a_{x,i+1/2}^+a_{x,i+1/2}^-\left[\mbox{\boldmath$q$}(\mbox{\boldmath$u$}_{L,i+1/2}^{n+1/2})
-\mbox{\boldmath$q$}(\mbox{\boldmath$u$}_{R,i+1/2}^{n+1/2})\right]}{a_{x,i+1/2}^+-a_{x,i+1/2}^-},
\end{eqnarray}
where the maximum and minimum wave speeds are defined by
\begin{equation}
a_{x,i+1/2}^+ = \mathrm{max}\{0,a_x^+(\mbox{\boldmath$u$}_{L,i+1/2}^{n+1/2}),a_x^+(\mbox{\boldmath$u$}_{R,i+1/2}^{n+1/2})\},
\end{equation}
\begin{equation}
a_{x,i+1/2}^- = \mathrm{min}\{0,a_x^-(\mbox{\boldmath$u$}_{L,i+1/2}^{n+1/2}),a_x^-(\mbox{\boldmath$u$}_{R,i+1/2}^{n+1/2})\}.
\end{equation}
Here $\mbox{\boldmath$u$}_{L,i+1/2}^{n+1/2}$ and $\mbox{\boldmath$u$}_{R,i+1/2}^{n+1/2}$
are the left and right state vectors of the primitive variables, which
are defined at the left and right edges of the cell interface,
respectively.
In the first order of spatial accuracy the left and right state vectors
reduce to $\mbox{\boldmath$u$}_{L,i+1/2}^{n+1/2} = \mbox{\boldmath$u$}_i^{n+1/2}$
and $\mbox{\boldmath$u$}_{R,i+1/2}^{n+1/2} = \mbox{\boldmath$u$}_{i+1}^{n+1/2}$.
For second-order accuracy in space the left and right state vectors
are interpolated as
\begin{equation}
\mbox{\boldmath$u$}_{L,i+1/2}^{n+1/2} = \mbox{\boldmath$u$}_i^{n+1/2}+\frac{1}{2}\Delta\mbox{\boldmath$u$}_i^n,~~
\mbox{\boldmath$u$}_{R,i+1/2}^{n+1/2} = \mbox{\boldmath$u$}_{i+1}^{n+1/2}-\frac{1}{2}\Delta\mbox{\boldmath$u$}_{i+1}^n,
\end{equation}
with the minmod limiter
\begin{equation}
\Delta\mbox{\boldmath$u$}_i^n = \frac{1}{2}
\left[\mathrm{sign}(\Delta^+\mbox{\boldmath$u$}_i^n)+\mathrm{sign}(\Delta^-\mbox{\boldmath$u$}_i^n)\right]
\mathrm{min}\{\left|\Delta^+\mbox{\boldmath$u$}_i^n\right|,\left|\Delta^-\mbox{\boldmath$u$}_i^n\right|\},
\end{equation}
or the monotonized central limiter
\begin{equation}
\Delta\mbox{\boldmath$u$}_i^n = \frac{1}{2}
\left[\mathrm{sign}(\Delta^+\mbox{\boldmath$u$}_i^n)+\mathrm{sign}(\Delta^-\mbox{\boldmath$u$}_i^n)\right]
\mathrm{min}\{2\left|\Delta^+\mbox{\boldmath$u$}_i^n\right|,2\left|\Delta^-\mbox{\boldmath$u$}_i^n\right|,
\frac{1}{2}\left|\Delta^+\mbox{\boldmath$u$}_i^n+\Delta^-\mbox{\boldmath$u$}_i^n\right|\},
\end{equation}
where $\Delta^+\mbox{\boldmath$u$}_i^n = \mbox{\boldmath$u$}_{i+1}^n-\mbox{\boldmath$u$}_i^n$
and $\Delta^-\mbox{\boldmath$u$}_i^n = \mbox{\boldmath$u$}_i^n-\mbox{\boldmath$u$}_{i-1}^n$.
In our formulation, the state vector of the primitive variables defined
at the half time step, $\mbox{\boldmath$u$}_i^{n+1/2}$, is computed from
a predictor step,
$\mbox{\boldmath$q$}_i^{n+1/2} = \mbox{\boldmath$q$}_i^n-(\Delta t^n/2\Delta x)
(\mbox{\boldmath$f$}_{x,i+1/2}^n-\mbox{\boldmath$f$}_{x,i-1/2}^n)$,
with the flux vector $\mbox{\boldmath$f$}_{x,i+1/2}^n$ calculated by
replacing the time step $n+1/2$ by the time step $n$ in equations (26) to (29).
This approach makes our code second order in time as well.

The time step is restricted by the Courant condition
$\Delta t^n = C_\mathrm{cour}\Delta x/\mathrm{max}\{|a_{x,i+1/2}^\pm|\}$
with $C_\mathrm{cour}<1$.
For multidimensional extensions, the numerical integration along the
$x$-direction is applied separately to the $y$- (and $z$-) directions
through the Strang-type dimensional splitting \citep{str68}.
In order to maintain second-order accuracy the order of
dimensional splitting is completely permuted in each successive sequence.

\section{Test Problems}

We have applied the numerical scheme with the general equation of state,
described in previous sections, to several test problems.
In all the test problems the minmod limiter and a Courant constant,
$C_\mathrm{cour} = 0.4$, are used.

\subsection{Relativistic Shock Tube}

The relativistic shock tube test is characterized initially by two
different states separated by a discontinuity.
As the initial discontinuity decays, distinct wave patterns
consisting of shock waves, contact discontinuities, and rarefraction
waves appear in the subsequent flow evolution.
In the relativistic shock tube problem the decay of the initial
discontinuity significantly depends on the tangential velocity since the
velocity components are coupled through the Lorentz factor in the
equations and the specific enthalpy also couples with the tangential
velocity \citep{pon00}.
As a result, the relativistic shock tube problem becomes more
challenging in the presence of tangential velocities.
This relativistic shock tube problem is very good test since it has an
analytic solution \citep{pon00,gia06} where the numerical solution can
be compared.

We have performed two sets of the two-dimensional relativistic shock
tube tests using the general equation of state for an electron-positron
gas.
The first test set is less relativistic, with $h\sim5$, while the second
test set is more severe, with a large initial internal energy ($h\gg1$).
For each of the test sets, two cases are presented, one having only
parallel velocity components while the other has, in addition, tangential
velocities.
In the first set, the initial left and right states for the case of only
parallel velocities are given by $\rho_L = 10$, $\rho_R = 1$, $v_{p,L} = 0$,
$v_{p,R} = 0$, $p_L = 13.3$, and $p_R = 10^{-6}$; for the case where
tangential velocities are included these additional initial values
are $v_{t,L} = 0.9$ and $v_{t,R} = 0.9$.
For the second set the initial left and right states for the case of
only parallel velocity are $\rho_L = 1$, $\rho_R = 1$, $v_{p,L} = 0$,
$v_{p,R} = 0$, $p_L = 10^3$, and $p_R = 10^{-2}$; now for the case
when tangential velocities are considered as well, $v_{t,L} = 0.99$ and
$v_{t,R} = 0.99$ are taken.
Here the subscripts $L$ and $R$ denote the left and right states
separated by an initial discontinuity placed along the main diagonal in
the two-dimensional computational plane.
Structures such as waves and discontinuities propagate along the
diagonal normal to the initial discontinuity.
Here $v_p$ is velocity parallel to the wave normal in the plane,
given by $v_p = \sqrt{v_x^2+v_y^2}$ and $v_t$ is velocity tangential to
the wave normal in the direction out of the plane, given by $v_t = v_z$.
The numerical computations are performed in a two-dimensional box with
$x = [0,1]$ and $y = [0,1]$ using $512\times512$ cells for the first set
and $2048\times2048$ cells for the second set.

\begin{figure}
\begin{center}
\includegraphics[scale=0.8]{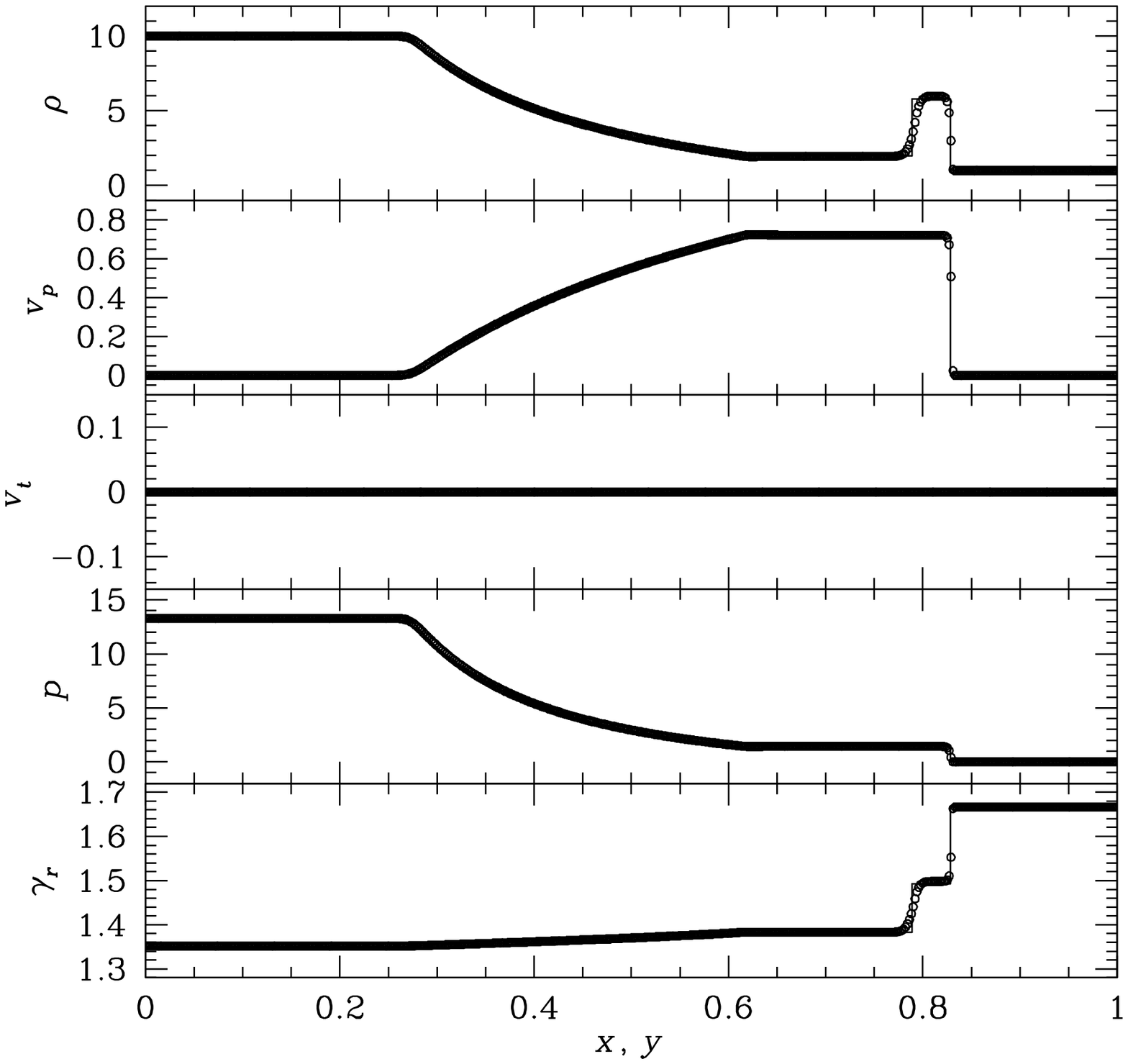}
\end{center}
\caption{(a) The two-dimensional relativistic shock tube test with the
general equation of state for an electron-positron gas for the case of
the parallel velocity in the first (moderately relativistic) test set.
The numerical computation is performed in a square box with $x = y = [0,1]$
using $512^2$ cells, and the wave structures are measured along the main
diagonal line $x = y = 0$ to $1$ at time $t = 0.4\sqrt{2}$.
The numerical solutions are marked with open circles and the analytical
solutions are plotted with solid lines.
(b) Same as in (a) but for the case of the included tangential velocity and at time
$t = 0.8\sqrt{2}$.}
\label{fig3}
\end{figure}

The results from the numerical computations carried with the general
equation of state for an electron-positron gas for the first test set
are shown in Figure \ref{fig3}.
Wave structures are measured along the main diagonal line $x = y = 0$ to
$1$ at times $t = 0.4\sqrt{2}$ and $0.8\sqrt{2}$ for the cases of the
only parallel and also tangential velocities, respectively.
The numerical solutions are marked with open circles and the analytical
solutions obtained using the numerical code available from \citet{gia06}
are plotted with solid lines.
Our numerical scheme with the proposed general equation of state is able
to reproduce all the wave structures with very good accuracy and stability, as
shown in Figure \ref{fig3}.
The shock waves and rarefraction waves are captured correctly, while the
contact discontinuities are relatively more smeared due to the use of the HLL
scheme and the minmod limiter.
The inclusion of the tangential velocity leads to the same basic wave pattern
as in the absence of tangential velocity, but the numerical
solutions are significantly modified.

Figure \ref{fig4} shows the results from the numerical computations with
the general equation of state for an electron-positron gas for the
second test set.
Structures are measured along the main diagonal line $x = y = 0$
to $1$ at times $t = 0.4\sqrt{2}$ and $1.8\sqrt{2}$ for the cases of the
parallel only and included tangential velocities, respectively.
The numerical solutions compare well with the
analytical solutions.
Shock waves and contact discontinuities propagate to the right, while
rarefraction waves move to the left.
As shown in Figure \ref{fig4}, all the wave structures are accurately
reproduced and their stability is good; however, the contact
discontinuities are somewhat smeared.
Again, the inclusion of the tangential velocity has a considerable
influence on the numerical solutions.

For a quantitative comparison with analytical solutions we have
calculated the norm errors of the rest mass density, parallel and
tangential velocities, and pressure defined by, e.g., for density,
$\|E(\rho)\|=\sum_{i,j}|\rho_{i,j}^N-\rho_{i,j}^A|\Delta x_i\Delta y_j$,
where the superscripts $N$ and $A$ represent numerical and analytical
solution, respectively.
The norm errors are given in Table \ref{tab1} and demonstrate a very
good agreement between the numerical and analytical solutions in all the
primitive variables.
We also carried out the two-dimensional relativistic shock tube tests
considered in Figures \ref{fig3} and \ref{fig4} for several other possible
combinations of the tangential velocity pairs $v_{t,L} = 0$, $0.9$, $0.99$
and $v_{t,R} = 0$, $0.9$, $0.99$.
In general, the stable wave structures are reproduced and the numerical
solutions are reasonably comparable to the analytical solutions for all
the different combinations of tangential velocity pairs.
As a consequence, the results from the two-dimensional relativistic
shock tube tests show that our code is able to robustly and
accurately capture discontinuities and waves.

\subsection{Relativistic Shock Reflection}

The relativistic shock reflection problem involves a collision between
two equal gas flows moving at relativistic speeds in opposite directions.
The collision of the two cold gases causes compression and heating of
the gases as kinetic energy is converted into internal energy.
This generates the two strong shock waves to propagate in opposite
directions, leaving the gas behind the shocks stationary.
The analytical solution of this relativistic shock reflection problem is
obtained by \citet{bla76} and \citet{gia06}.

We have tested the two-dimensional relativistic shock reflection problem
using the general equation of state for an electron-positron gas.
Two cases are presented here.
As for the shock tube tests,
one includes only the parallel velocity and the other includes in
addition the tangential velocity.
The left and right states for the case of only parallel velocity are initially
given by $\rho_L = 1$, $\rho_R = 1$, $v_{p,L} = 0.99$, $v_{p,R} = -0.99$,
$p_L = 10^{-6}$, and $p_R = 10^{-6}$, and for the case when the tangential
velocity is included, oppositely directed tangential velocities $v_{t,L} = 0.1$
and $v_{t,R} = -0.1$ are assumed to be present on either side of the plane.
Here the subscripts $L$ and $R$ stand for the left and right states
separated by the initial collision points located along the main diagonal
in the two-dimensional computational plane.
The two shock waves propagate diagonally in opposite directions, and should
keep symmetric with respect to the initial collision points.
As in the relativistic shock tube tests, $v_p$ is velocity parallel to
the wave normal in the plane and $v_t$ is velocity tangential to the wave
normal in the direction out of the plane.
The numerical computations are carried out in a two-dimensional box with
$x = [0,1]$ and $y = [0,1]$ using $512\times512$ cells.

Figure \ref{fig5} shows the results from our relativistic shock reflection
tests with the general equation of state for an electron-positron gas.
Structures are measured along the main diagonal line $x = y = 0$ to $1$
at time $t = 0.8\sqrt{2}$.
The numerical solutions are in very good agreement with the analytical
solutions.
In both cases, the shock wave is
resolved by two numerical cells, and there are no numerical oscillations
behind the shocks.
As shown in Figure \ref{fig5}, the compression ratio between shocked and
unshocked gases is about $30$ for the rest mass density and about $70$ for
the pressure in the case of only parallel velocities; the inclusion of the
tangential velocities increases the compression ratio to about $40$ for
the rest mass density and to about $140$ for the pressure.
Near $x = y = 0.5$, the density distribution slightly underestimates the
analytical solution and a stationary discontinuity in the tangential
velocity is somewhat diffused due to the numerical effect of reflection
heating phenomena.
The norm errors of the rest mass density, parallel and tangential
velocities, and pressure are also given in Table \ref{tab1}.
A direct comparison with the analytical solutions shows that the
measured errors are very small for all the primitive variables.

The accuracy of numerical solutions depends on the number of cells
spanned by the computational box.
We have run the relativistic shock reflection test in Figure \ref{fig5}(b)
with different numerical resolutions to check the convergence rate.
Except for the numerical resolutions the initial conditions are identical
to those used in the test in Figure \ref{fig5}(b).
We have computed the norm errors for rest mass density, velocities, and
pressure with different resolutions.
Numerical resolutions of $16^2$, $32^2$, $64^2$, $128^2$, $256^2$,
and $512^2$ cells give norm errors for the rest mass density of $7.56$,
$4.74$, $2.38$, $1.17$, $0.48$, and $0.25$, respectively.
As expected for discontinuous problems, first-order convergence in the
norm errors for rest mass density is obtained with increasing the
numerical resolution.
Similar clear trends toward convergence are seen in the norm errors for
velocities and pressure.

\subsection{Relativistic Blast Wave}

In a relativistic blast wave test a large amount of energy is initially
deposited in a small finite spherical volume and the subsequent
expansion of that overpressured region is evolved forward in time.
This produces a spherical shock propagating outward from an initial
discontinuity at an arbitrary radius.
This radial blast wave explosion provides a useful test problem to
explore the spherically symmetric properties in highly relativistic flow
speeds.

We have performed the three-dimensional relativistic blast wave test
with the general equation of state for an electron-proton gas.
The initial condition for the relativistic blast wave problem consists
of two constant states given by $\rho_I = 1$, $\rho_O = 1$, $v_I = 0$,
$v_O = 0$, $p_I = 10^3$, and $p_O = 1$, where subscripts $I$ and $O$
represent the inner and outer states separated by an initial
discontinuity at the radius $r = 0.5$ in the three-dimensional
computational box.
The numerical computations are performed in a three-dimensional box with
$x = [0,1]$, $y = [0,1]$, and $z = [0,1]$ using $256\times256\times256$
cells.
Outflow boundary conditions are used at all boundaries except at the
symmetry axis where reflecting boundary conditions are imposed.

Figure \ref{fig6} shows the profiles of rest mass density, radial
velocity, and pressure along the radial distance in the relativistic
blast wave tests.
Radial structures are measured along the main diagonal line
$x = y = z = 0$ to $1$ at time $t = 0.4$.
Numerical computations carried out with our general equation of state
for an electron-proton gas, as well as with the ideal gas equation of
state with constant adiabatic index $\gamma = 5/3$ are shown.
In both cases, high Lorentz factors (up to $\sim15$) are generated and
the radial symmetry is well preserved.
However, compared to the $\gamma = 5/3$ case, the results obtained with
the electron-proton case show significant differences.
In fact, the spherical shock wave has a higher density peak and propagates
at a slower radial velocity in the electron-proton case.

In Figure \ref{fig7}, we present the images of rest mass density,
pressure, and Lorentz factor in the relativistic blast wave test with
the general equation of state for an electron-proton gas.
The images are shown in logarithmic scales on the plane $x = 0$ at time
$t = 0.4$.
As shown in the images, the spherical shock wave propagates to larger
radius very well, preserving the initial spherical symmetry.

\section{Relativistic Axisymmetric Jets}

As a practical astrophysical problem using this code, we consider the
propagation of relativistic axisymmetric jets in three dimensions.
The simulation of a relativistic jet has been presented as a test
simulation in almost all relativistic hydrodynamic codes, using the
ideal gas equation of state with a constant adiabatic index
\citep[e.g.,][]{alo99,del02,mig05a}.
We have performed this relativistic jet simulation to confirm the
accuracy and robustness of our numerical scheme incorporating the
proposed general equation of state as well as to make a preliminary
investigation of the influence of the equation of state on the
propagation of relativistic jets.

Some previous investigations of relativistic jet propagation have
concentrated on the importance of a realistic equation of state in
relativistic jet flows.
\citet{mig07} addressed the effect of a realistic equation of state on
relativistic (magnetized) flows including relativistic jets.
They showed that the choice of a realistic equation of state can
significantly alter the solution when large temperature gradients are
present.
To study the evolution of low-power jets \citet{per07} performed
numerical simulations using an equation of state for a two-component
relativistic gas that separately treats leptonic and baryonic matter.
\citet{ros08} examined the dynamical evolution of relativistic light
jets in the presence of an induced perturbation using a Synge-like
equation of state for a single-component relativistic gas.

The approximate propagation velocity of the jet through the homogeneous
ambient medium can be derived from momentum flux balance between the
beam and ambient medium in the reference frame of the working surface
that separates beam and shocked ambient gas \citep[e.g.,][]{mar97}.
Assuming pressure equilibrium between the beam and the ambient medium,
the one-dimensional jet advance velocity, estimated in the rest frame of
the ambient medium, is then
\begin{equation}
v_a = \frac{\Gamma_b\sqrt{\eta h_b/h_a}}{1+\Gamma_b\sqrt{\eta h_b/h_a}}v_b,
\end{equation}
where $\eta$ and $\Gamma_b$ are given by $\eta = \rho_b/\rho_a$ and
$\Gamma_b = 1/\sqrt{1-v_b^2}$.
Here the subscripts $b$ and $a$ indicate the beam and the ambient medium,
respectively.
The morphology and dynamics of the relativistic jet propagating into the
homogeneous medium is commonly specified by the beam to ambient medium
density ratio $\eta$, the beam Lorentz factor $\Gamma_b$, and the beam
Mach number $M_b = v_b/c_{s,b}$.

We have considered the three-dimensional simulations of relativistic
jets using different equations of state.
Table \ref{tab2} lists the simulation parameters of the four different
cases corresponding to the ideal gas equation of state with constant
adiabatic index $\gamma = 5/3$ and $4/3$ and the proposed general
equation of state for electron-positron and electron-proton gases.
In all these cases, we choose the beam density $\rho_b = 0.1$, the ambient
medium density $\rho_a = 10$, the beam velocity $v_b = 0.99$ and we
assume that the beam is in pressure equilibrium with the ambient medium,
$p_b = p_a$.
This gives the density contrast $\eta = 10^{-2}$, the beam Lorentz
factor $\Gamma_b = 7.1$, and the beam Mach number $M_b = 2$.
The numerical simulations have been performed in the three-dimensional
computational box with $x = [0,1]$, $y = [0,1]$, and $z = [0,4]$ using a
uniform numerical grid of $128\times128\times512$ cells.
The beam has an initial radius $r_b = 1/8$ (corresponding to $16$ cells), is
launched from the origin, and propagates through a uniform static
ambient medium along the positive $z$-direction.
Outflow boundary conditions are set at all boundaries except along the
symmetry axis where reflecting boundary conditions are used and in the
injection region where an inflow boundary condition is imposed to keep
the beam constantly fed.
The monotonized central limiter and the Courant constant
$C_\mathrm{cour} = 0.2$ are used in all these jet propagation cases.

The images in Figure \ref{fig8} display the logarithms of the rest mass
density on the plane $x = 0$ at time $t = 6$ for the four different
cases in the simulations of relativistic axisymmetric jets.
The bow shock, the beam, and the cocoon surrounding the beam can be
clearly identified in all four cases, confirming the ability of our code
to follow complex relativistic flows.
For each case, a bow shock that separates the shocked jet material from the
shocked ambient medium is driven into the ambient medium.
The beam itself is terminated by a Mach disk where
much of the beam's kinetic energy is converted into internal energy.
Shocked jet material flows backward along the working surface into a
cocoon, resulting in the development and mixture of turbulent vortices
in the cocoon, and the interaction of these vortices with the beam forms
oblique internal shocks within the beam close to the Mach disk, which
causes the deceleration of the jet.
The four different cases, however, show differences in specific
morphological and dynamical properties.
The cold jet (case A) propagates at the slowest velocity, and the
jet produces a broad bow shock, a thick cocoon, and has the Mach disk located
quite far behind the bow shock.
On the contrary, the hot jet (case B) is dominated by a narrow bow
shock, has a thin cocoon, and its Mach disk lies very close to the bow shock,
thanks to its having the fastest advance velocity.
The electron-positron and the electron-proton jets (cases C and D)
propagate faster than the cold jet, but slower than the hot jet.
Therefore, the electron-positron and the electron-proton jets
possess morphological and dynamical properties intermediate between the
cold and the hot jets.
In terms of our simulation parameters, the electron-positron jet tends to be more
similar to the cold jet, while the electron-proton jet seems to share
more features with the hot jet.

In Figure \ref{fig9}, the position of the bow shock is plotted as a
function of time for the four different gases.
The symbols mark the numerical estimate for a selected time interval for
each case, and the lines represent the one-dimensional theoretical
estimate in equation (32).
The numerical simulations are in good agreement with the one-dimensional
theoretical estimates for the bow shock location for all cases.

\section{Conclusion}

Most numerical codes for special relativistic hydrodynamics have
used the ideal gas equation of state with a constant adiabatic
index, but this is a poor approximation for most relativistic
astrophysical flows.
We proposed a new general equation of state for a multi-component
relativistic gas, based on the Synge equation of state for a relativistic
perfect gas \citep{syn57}.
Our proposed general equation of state is very efficient and suitable
for numerical relativistic hydrodynamics since it has an analytic
expression.
The thermodynamic quantities computed using the proposed general
equation of state behave correctly asymptotically in the limits of hot and
cold gases.
For intermediate regimes the thermodynamic quantities
vary between those for the two limiting cases, depending on the composition of the
relativistic gas.

We also presented a multidimensional relativistic hydrodynamics code
incorporating the proposed general equation of state for a multi-component
relativistic gas.
Our numerical code is based on the HLL scheme \citep{har83}, which
avoids a full
characteristic decomposition of the relativistic hydrodynamic equations
and uses an approximate solution to the Riemann problem for flux
calculations.
Since the numerical code is fully explicit, retaining a second-order
accuracy in space and time, it is simple to extend the code to
different geometries or to produce parallelized versions of this code.
The analytical formulation of the proposed equation of state and the
numerical scheme being free of complete characteristic decomposition
make the code very
efficient and robust in ultrarelativistic multidimensional problems.

The accuracy and robustness of the code are demonstrated in two
dimensions using the test problems of the relativistic shock tube and
the relativistic shock reflection and in three dimensions using the test
problem of the relativistic blast wave.
The direct comparisons of numerical results with analytical solutions
show that shocks and discontinuities are correctly resolved even in
highly relativistic test problems with nonvanishing tangential
velocities.
Results from the three-dimensional simulations of the relativistic
axisymmetric jets demonstrate the ability of our code to follow complex
relativistic flows as well as the flexibility enough to be applied to
practical astrophysical problems.
These simulations show that the morphology and dynamics of the
relativistic jets are significantly influenced by the different equation
of state and by different compositions of a relativistic perfect gas.

\acknowledgments

This work was supported by the National Research Foundation of Korea
Grant funded by the Korean Government (NRF-2009-351-C00029).

\clearpage

\clearpage

\clearpage

\clearpage

\begin{figure}
\begin{center}
\includegraphics[scale=0.8]{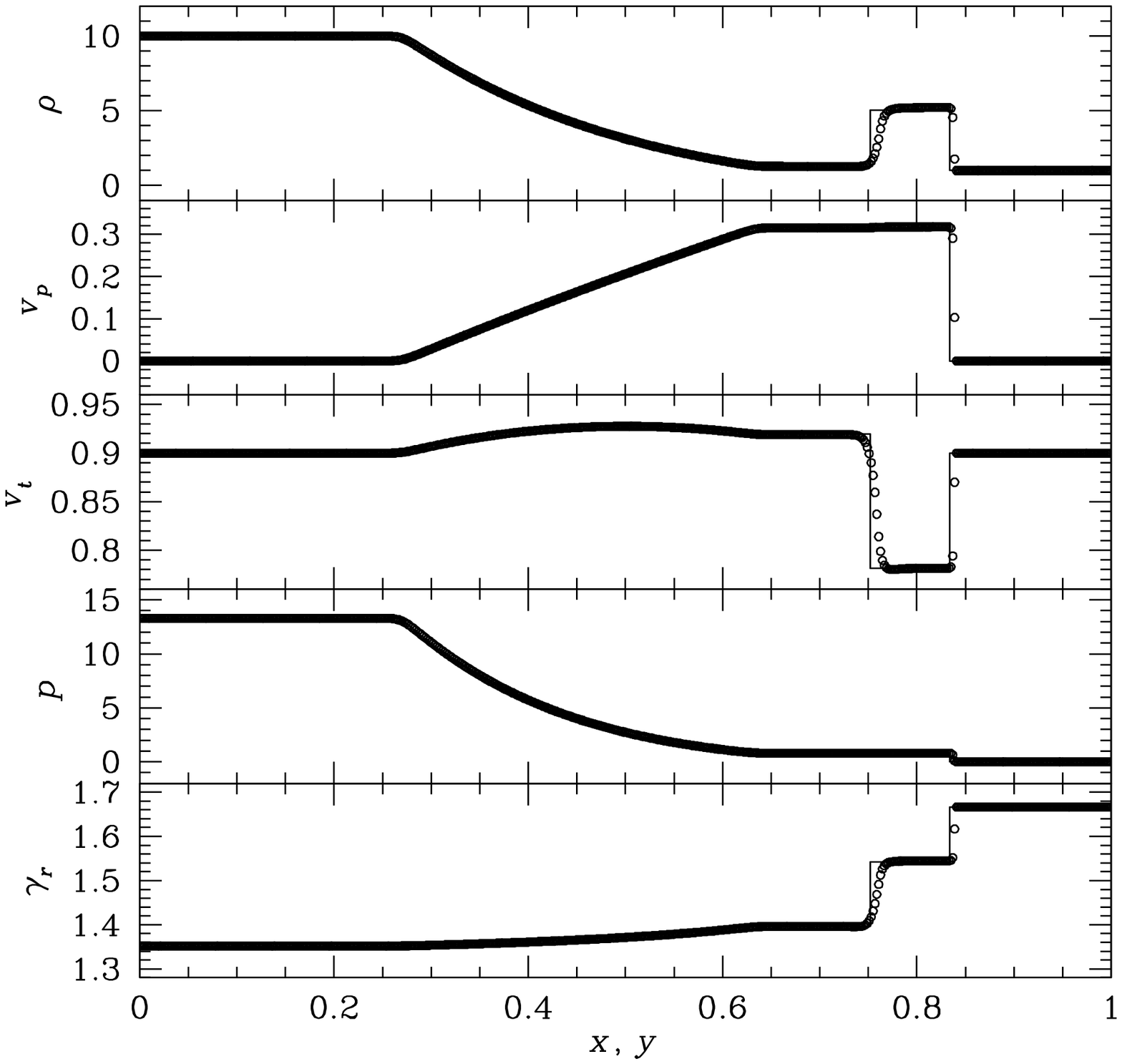}
\end{center}
\end{figure}

\clearpage

\begin{figure}
\begin{center}
\includegraphics[scale=0.8]{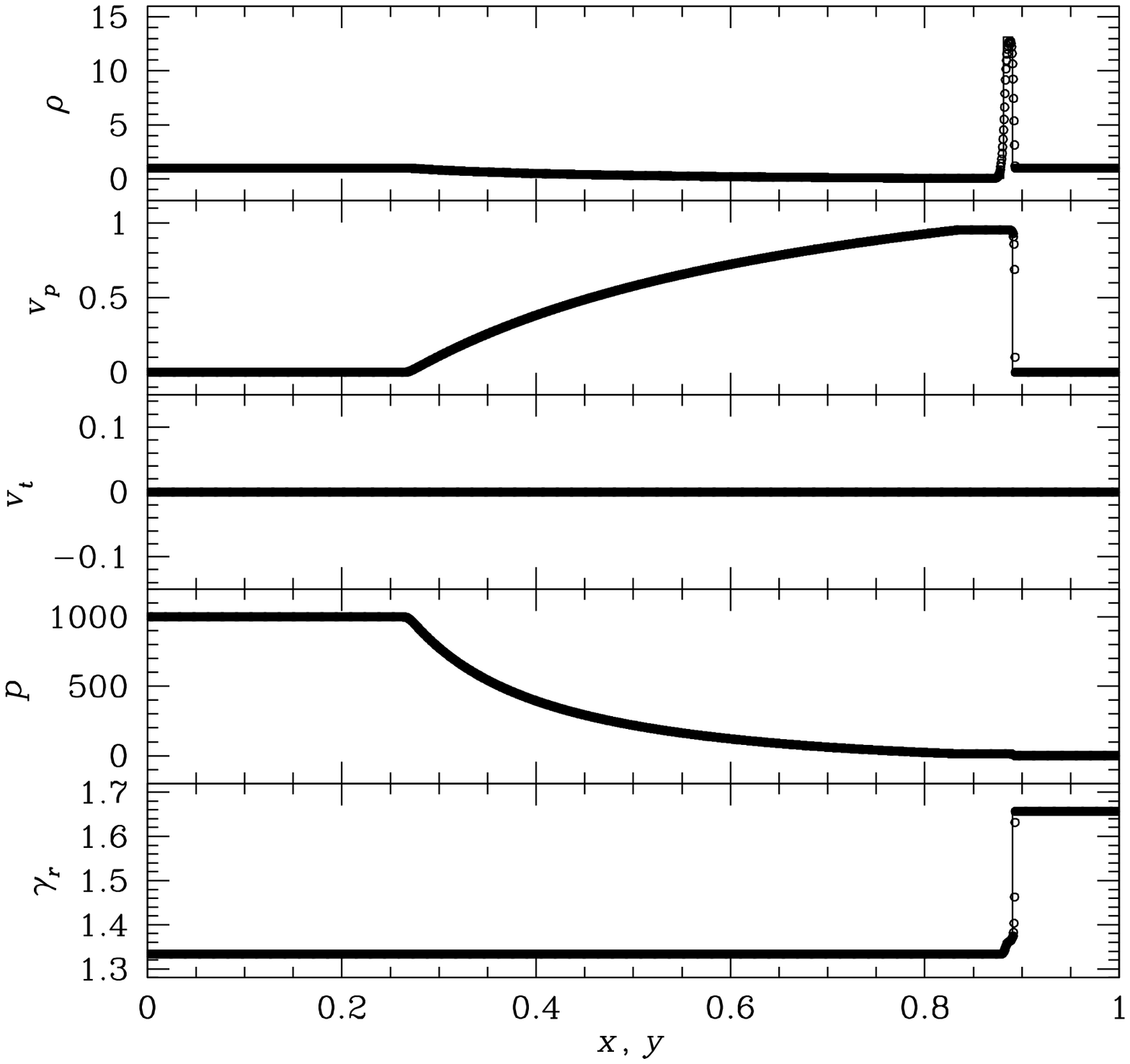}
\end{center}
\caption{(a) The two-dimensional relativistic shock tube test with the
general equation of state for an electron-positron gas for the case of
the parallel velocity in the second (highly relativistic) test set.
The numerical computation is performed in a square box with $x = y = [0,1]$
using $2048^2$ cells, and the wave structures are measured along the main
diagonal line $x = y = 0$ to $1$ at time $t = 0.4\sqrt{2}$.
The numerical solutions are marked with open circles and the analytical
solutions are plotted with solid lines.
(b) Same as in (a) but for the case of the included tangential velocity and at time
$t = 1.8\sqrt{2}$.}
\label{fig4}
\end{figure}

\clearpage

\begin{figure}
\begin{center}
\includegraphics[scale=0.8]{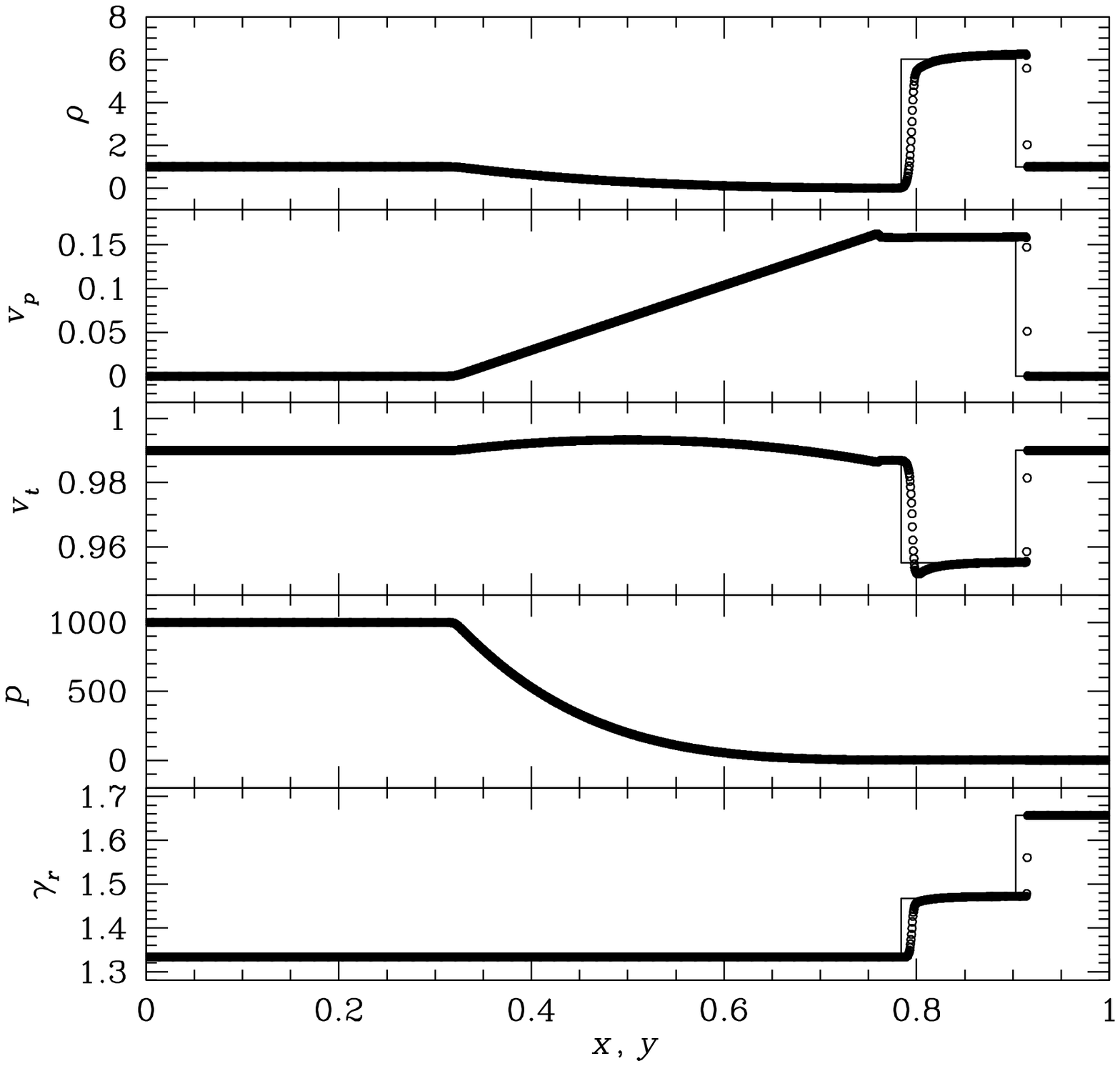}
\end{center}
\end{figure}

\clearpage

\begin{figure}
\begin{center}
\includegraphics[scale=0.8]{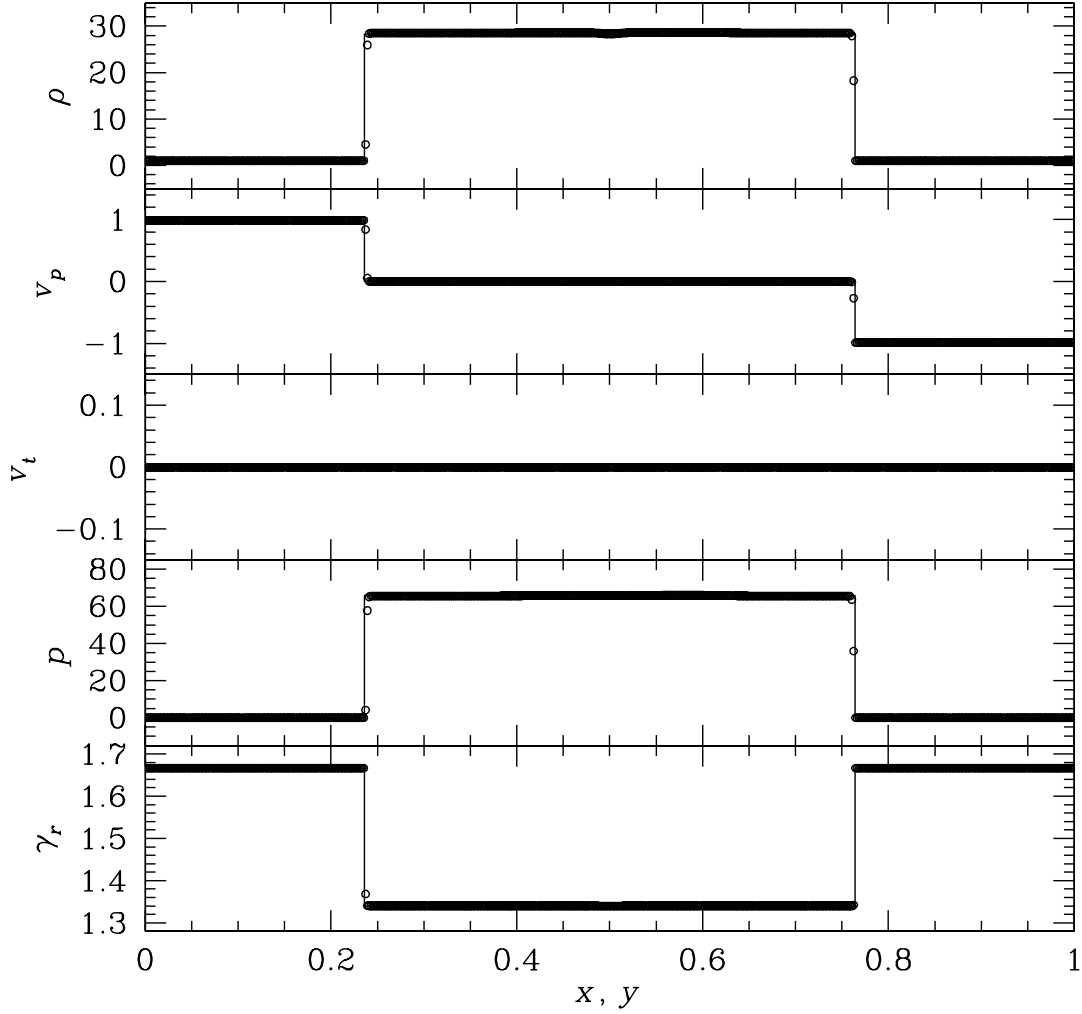}
\end{center}
\caption{(a) The two-dimensional relativistic shock reflection tests
with the general equation of state for an electron-positron gas for the
case of the parallel velocity.
The numerical computations are performed in a square box with $x = y = [0,1]$
using $512^2$ cells, and structures are measured along the main diagonal
line $x = y = 0$ to $1$ at time $t = 0.8\sqrt{2}$.
The numerical solutions are marked with open circles and analytical
solutions are plotted with solid lines.
(b) Same as in (a) but for the case of the included tangential velocity.}
\label{fig5}
\end{figure}

\clearpage

\begin{figure}
\begin{center}
\includegraphics[scale=0.8]{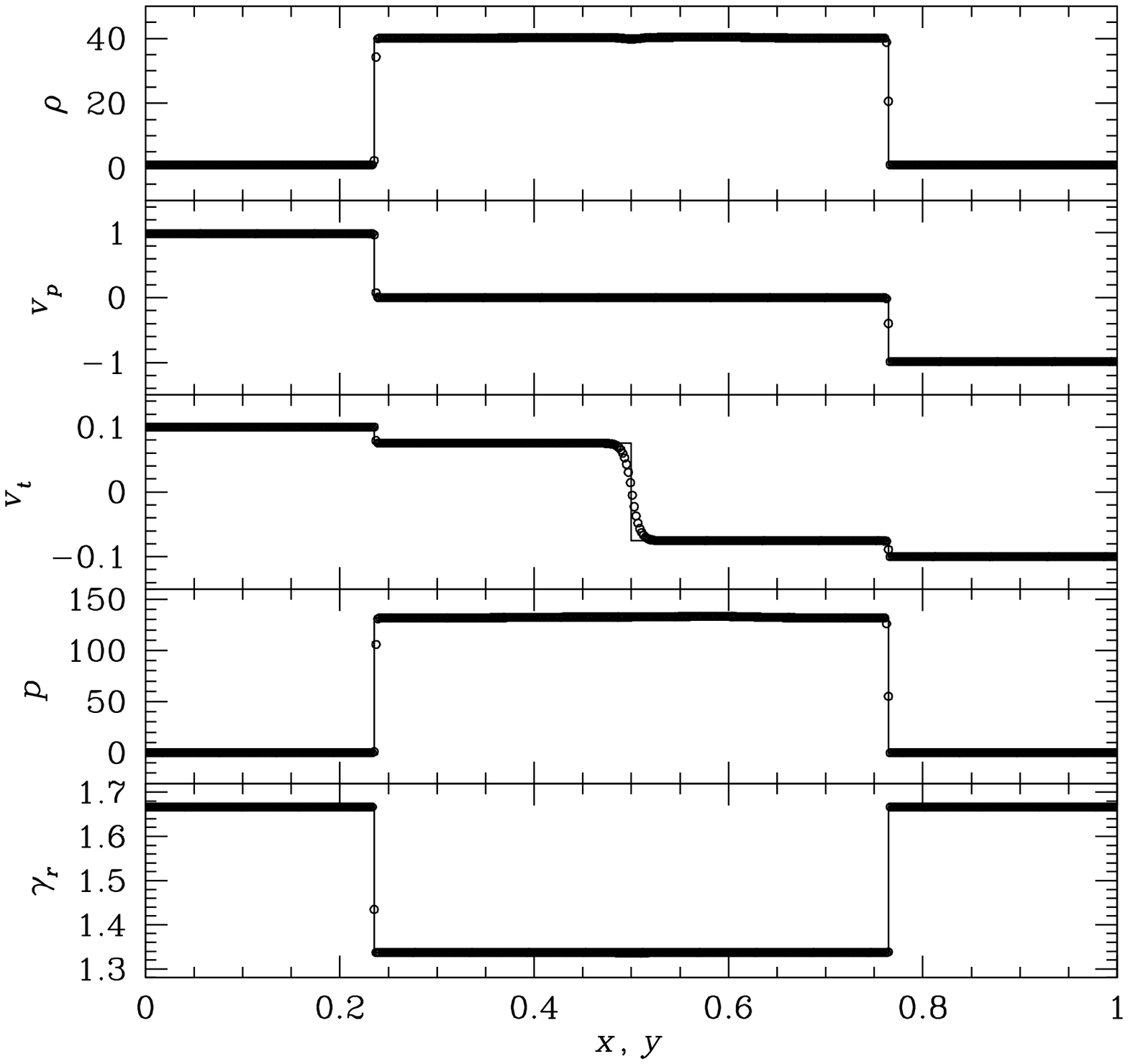}
\end{center}
\end{figure}

\clearpage

\begin{figure}
\begin{center}
\includegraphics[scale=0.8]{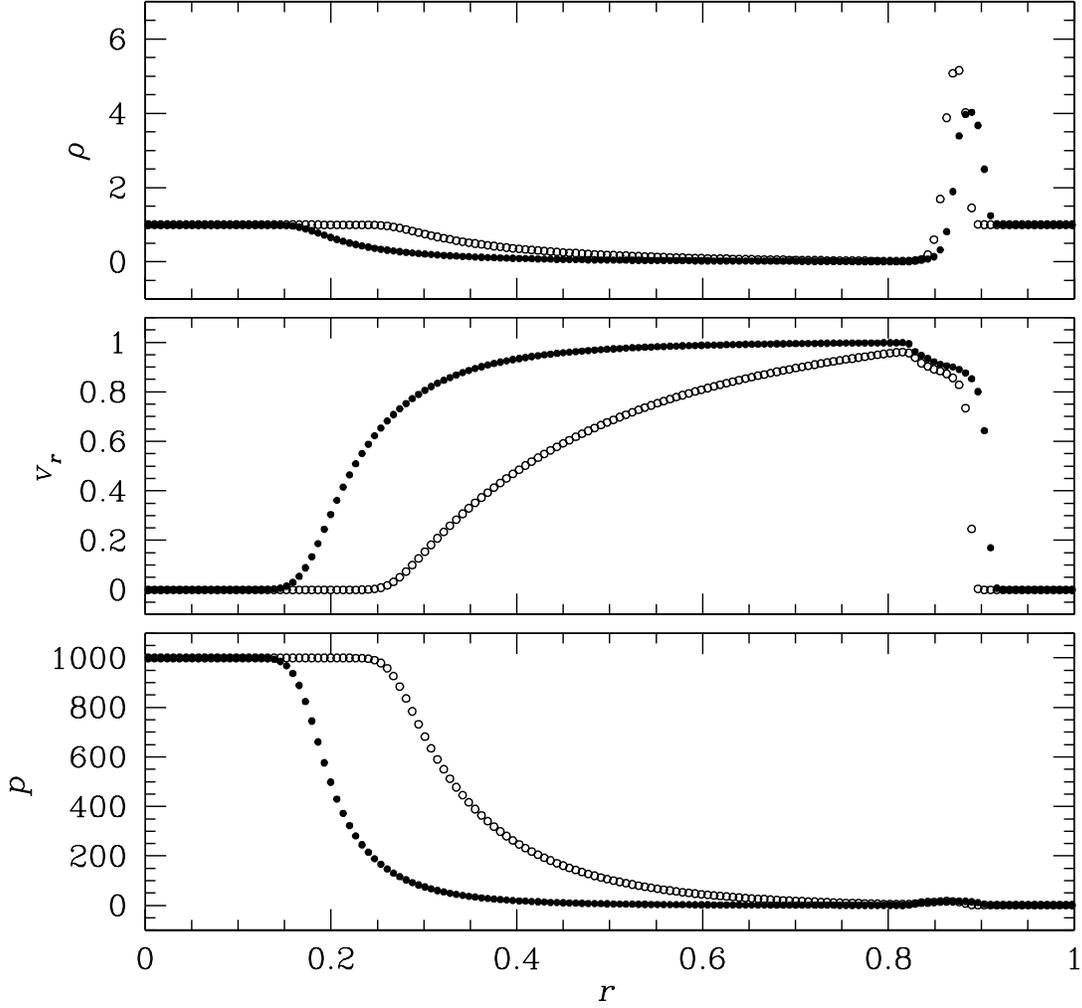}
\end{center}
\caption{Profiles of rest mass density $\rho$, radial velocity $v_r$,
and pressure $p$ along the radial distance $r$ in the three-dimensional
relativistic blast wave tests.
The numerical computations are performed in a cube box with $x = y = z = [0,1]$
using $256^3$ cells, and radial structures are measured along the main
diagonal line $x = y = z = 0$ to $1$ at time $t = 0.4$.
Numerical computations carried out with the general equation of state for an
electron-proton gas and those done with an ideal gas equation of state with constant
adiabatic index of $\gamma = 5/3$ are marked using open and filled circles,
respectively.}
\label{fig6}
\end{figure}

\clearpage

\begin{figure}
\begin{center}
\includegraphics[scale=0.8]{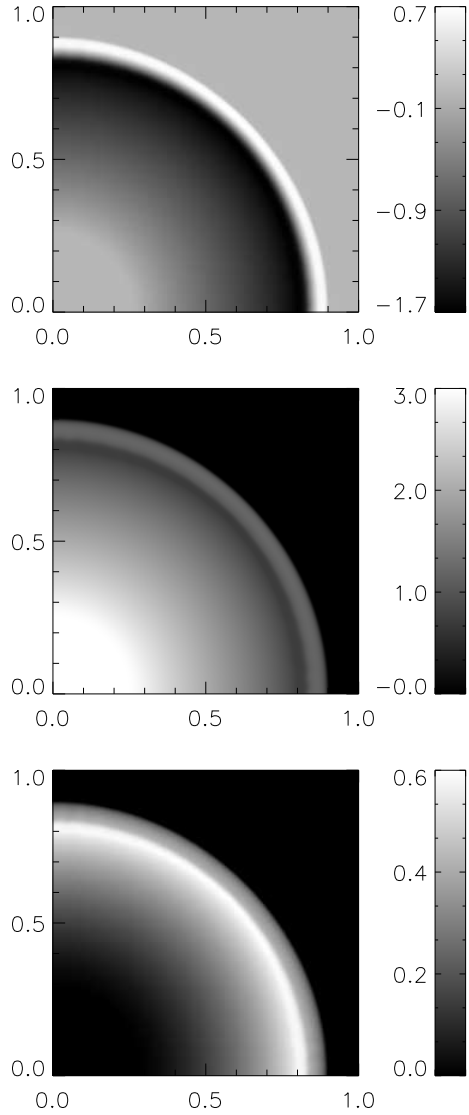}
\end{center}
\caption{Images of rest mass density (top), pressure (middle), and
Lorentz factor (bottom) in the three-dimensional relativistic blast wave
test with the general equation of state for an electron-proton gas.
The numerical computations are performed in a cube box with $x = y = z = [0,1]$
using $256^3$ cells, and the images are shown in logarithmic scales on
the plane $x = 0$ at time $t = 0.4$.}
\label{fig7}
\end{figure}

\clearpage

\begin{figure}
\begin{center}
\includegraphics[scale=0.8]{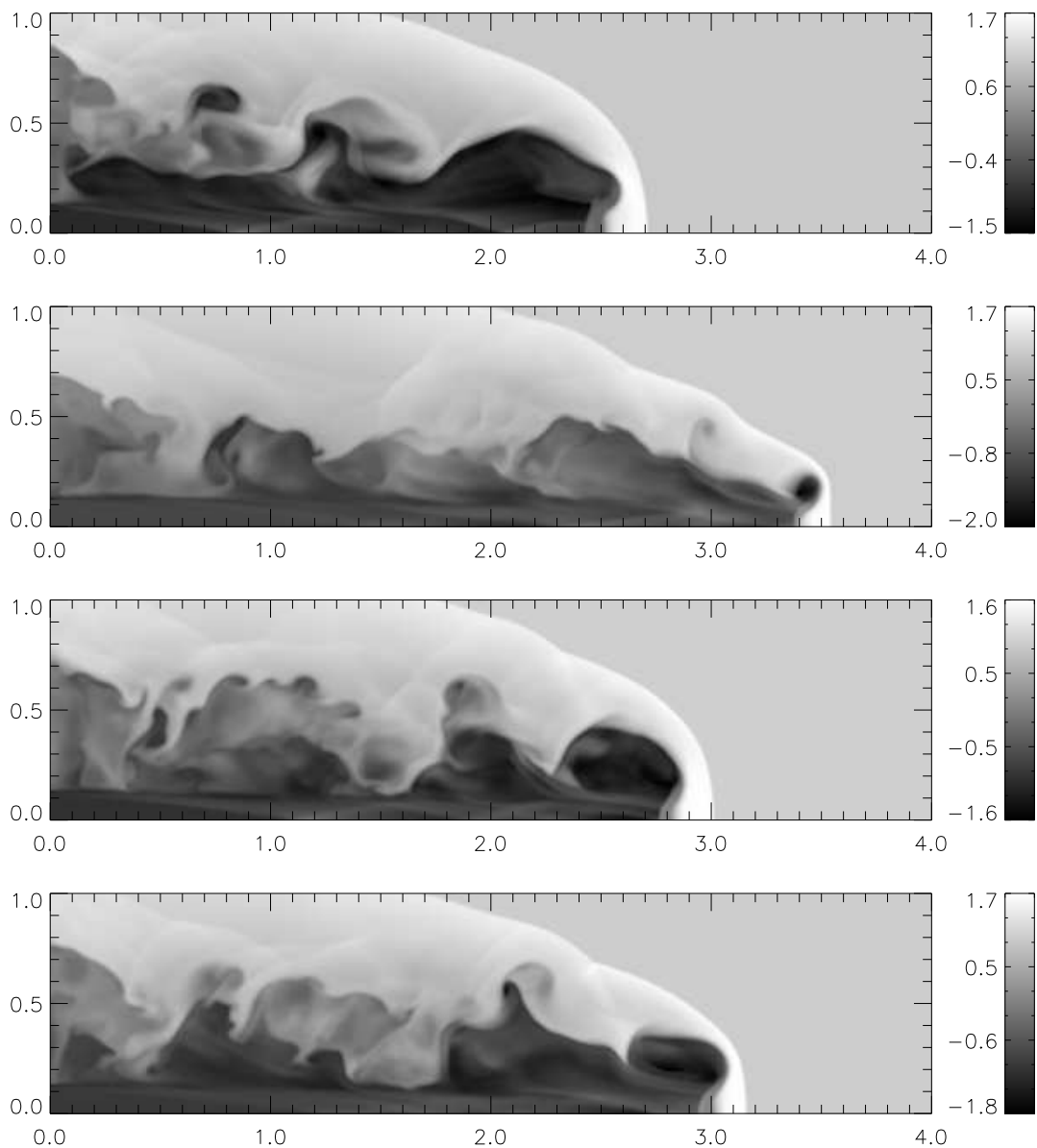}
\end{center}
\caption{Images of the rest mass density for cases A to D (top to bottom)
in the three-dimensional simulations of relativistic axisymmetric jets.
The numerical simulations have been performed in the computational box
with $x = [0,1]$, $y = [0,1]$, and $z = [0,4]$ using $128\times128\times512$
cells, and the images are shown in logarithmic scales on the plane $x = 0$
at time $t = 6$.}
\label{fig8}
\end{figure}

\clearpage

\begin{figure}
\begin{center}
\includegraphics[scale=0.8]{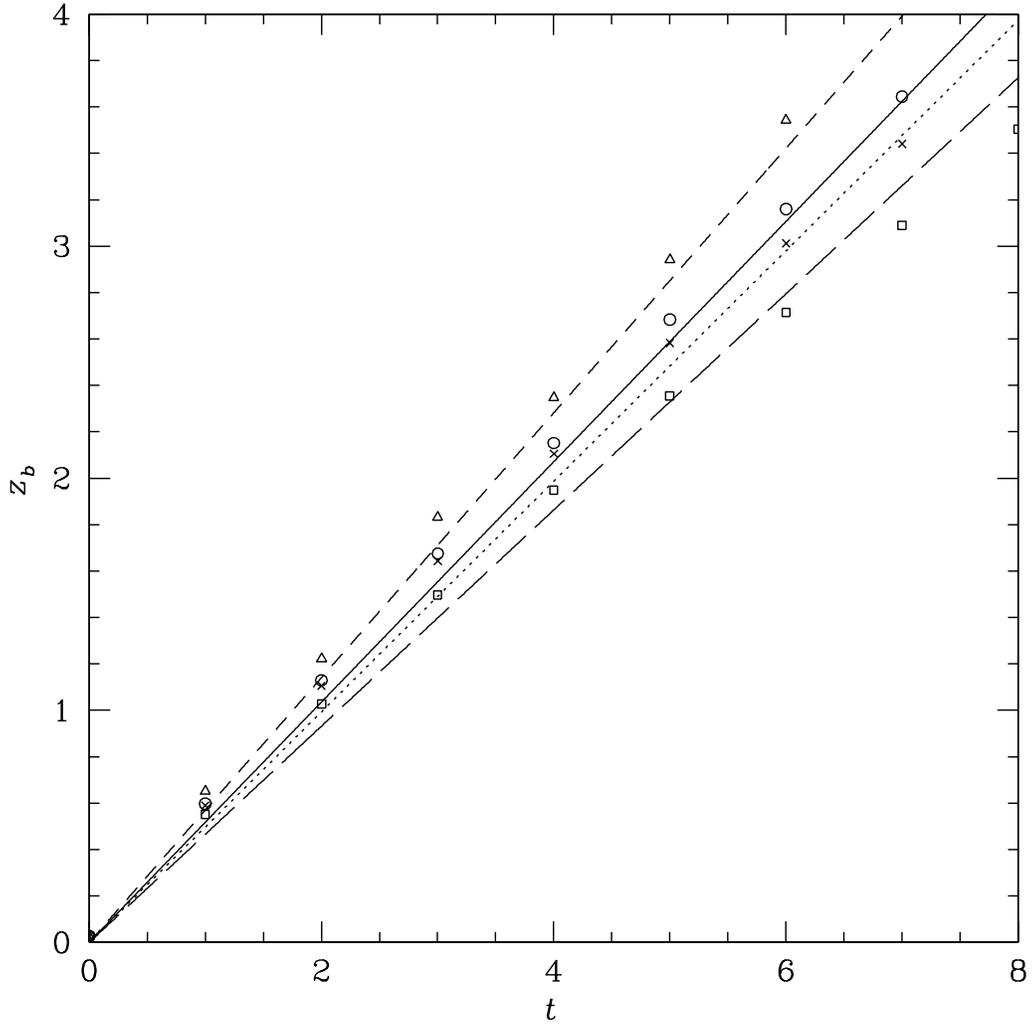}
\end{center}
\caption{The position of the bow shock as a function of time for cases
A to D in the three-dimensional simulations of relativistic axisymmetric
jets.
The numerical results are marked with squares, triangles, crosses, and
circles for cases A to D at selected time intervals, respectively, and
the long dashed, short dashed, dotted, and solid lines give the corresponding
one-dimensional theoretical estimates.}
\label{fig9}
\end{figure}

\clearpage

\begin{deluxetable}{ccccc}
\tablewidth{0pt}
\tablecaption{Norm Errors for Relativistic Shock Tube/Reflection Tests
\label{tab1}}
\tablehead{
\colhead{Test} &
\colhead{$\|E(\rho)\|$} &
\colhead{$\|E(v_p)\|$} &
\colhead{$\|E(v_t)\|$} &
\colhead{$\|E(p)\|$}}
\startdata
 RST3a & 4.96E$-$02 & 3.91E$-$03 & 0.00E$+$00 & 3.46E$-$02 \\
 RST3b & 7.85E$-$02 & 2.38E$-$03 & 1.64E$-$03 & 3.11E$-$02 \\
 RST4a & 4.43E$-$02 & 2.83E$-$03 & 0.00E$+$00 & 5.12E$-$01 \\
 RST4b & 1.43E$-$01 & 2.20E$-$03 & 8.46E$-$04 & 4.21E$-$01 \\
 RSR5a & 1.81E$-$01 & 2.33E$-$03 & 0.00E$+$00 & 2.61E$-$01 \\
 RSR5b & 2.48E$-$01 & 2.92E$-$03 & 9.92E$-$04 & 6.91E$-$01 \\
\enddata
\tablecomments{The test designation indicates relativistic shock tube
(RST) and relativistic shock reflection (RSR) problems followed by
corresponding figure numbers (3 to 5) and labels (a and b).}
\end{deluxetable}

\clearpage

\begin{deluxetable}{ccccccccc}
\tablewidth{0pt}
\tablecaption{Simulation Parameters for Relativistic Axisymmetric Jets
\label{tab2}}
\tablehead{
\colhead{Case} &
\colhead{$\eta$} &
\colhead{$\Gamma_b$} &
\colhead{$M_b$} &
\colhead{$p_b$} &
\colhead{EOS} &
\colhead{$\chi$} &
\colhead{$N_\mathrm{cell}$} &
\colhead{$t$}}
\startdata
 A & $10^{-2}$ & 7.1 & 2 & $2.32\times10^{-2}$ & $5/3$    & -   & $128\times128\times512$ & 8 \\
 B & $10^{-2}$ & 7.1 & 2 & $6.94\times10^{-2}$ & $4/3$    & -   & $128\times128\times512$ & 6 \\
 C & $10^{-2}$ & 7.1 & 2 & $3.48\times10^{-2}$ & $e^-e^+$ & 0.0 & $128\times128\times512$ & 7 \\
 D & $10^{-2}$ & 7.1 & 2 & $4.23\times10^{-2}$ & $e^-p^+$ & 1.0 & $128\times128\times512$ & 7 \\
\enddata
\tablecomments{Here $\eta$ is the density ratio of the beam to the
ambient medium, $\Gamma_b$ is the beam Lorentz factor, $M_b$ is the beam
Mach number, $p_b$ is the uniform pressure in both the beam and the ambient
medium, EOS is the type of equation of state, $\chi$ is the relative
fraction of proton and electron number densities, $N_\mathrm{cell}$ is
the numerical resolution of the three-dimensional computational box, and
$t$ is the time out to which each simulation is followed.}
\end{deluxetable}

\end{document}